\documentclass{PoS}
\usepackage{subcaption}
\usepackage{graphicx}
\usepackage{mathtools}
\usepackage{cite}

\title{Measurement of the inelastic proton-proton cross section at $\sqrt{S}$ = 13 TeV}
\ShortTitle{Measurement of the inelastic proton-proton cross section at $\sqrt{S}$ = 13 TeV}

\author{\speaker{H. Van Haevermaet} for the CMS Collaboration\\
	University of Antwerp, Belgium\\
	E-mail: \email{hans.vanhaevermaet@uantwerpen.be}}

\abstract{A measurement of the inelastic proton-proton cross section at a centre-of-mass energy of $\sqrt{s}$ = 13 TeV is presented. The analysis is performed using the CMS detector, in particular with information from forward calorimetry at pseudorapidities of 3.0 < $\eta$ < 5.2 and -6.6 < $\eta$ < -3.0. A visible cross section is measured in two different detector acceptances and finally extrapolated to the full inelastic phase space domain. The results are compared with those of other experiments, and with models used to describe high-energy hadronic interactions. }

\FullConference{XXIV International Workshop on Deep-Inelastic Scattering and Related Subjects\\
		11-15 April, 2016\\
		DESY Hamburg, Germany}

\begin{document}

\section{Introduction}
The hadronic cross section is a fundamental observable in high energy particle physics, and has been measured in many experiments, covering several orders of magnitude in centre-of-mass energy. It can be decomposed in an elastic and inelastic part, where the latter can again be decomposed into diffractive and non-diffractive contributions that can be described by non-perturbative, phenomenological models. However, these models have large uncertainties when one wants to extrapolate existing measurements towards higher centre-of-mass energies. Therefore precise measurements of the hadronic cross sections are needed to provide input to phenomenological models and for the tuning of Monte Carlo event generators. In addition, the value of the inelastic proton-proton cross section ($\sigma_{\text{inel}}$) is used to estimate the average pile-up in the data, i.e. the number of simultaneous proton-proton interactions occurring in the same bunch crossings of the accelerator beams. This can be an important background to new physics searches and the measurement of $\sigma_{\text{inel}}$ is thus important to control this pile-up contribution in proton-proton interactions measured at the LHC.

Experiments at the LHC measured $\sigma_{\text{inel}}$ at a centre-of-mass energy of $\sqrt{s}$ = 7 TeV, reporting values ranging between 66.9 mb and 73.2 mb \cite{Abelev:2012sea, Aad:2014dca, CMS:2011xpa, Chatrchyan:2012nj, Aaij:2014vfa, Antchev:2013iaa}. This large spread is mainly due to the different methods used to extrapolate the measured value to the total inelastic phase space, and indicates that a precise determination of $\sigma_{\text{inel}}$ is very challenging. The most precise values of $\sigma_{\text{inel}}$ are currently given by the ATLAS Collaboration \cite{Aad:2014dca} : $\sigma_{\text{inel}} = 71.34 \pm 0.9$ mb; and by the TOTEM  Collaboration \cite{Antchev:2013iaa}: $\sigma_{\text{inel}} = 72.9 \pm 1.5$ mb. In addition, the TOTEM Collaboration also performed a measurement of $\sigma_{\text{inel}}$ at a centre-of-mass energy of $\sqrt{s}$ = 8 TeV and reported a value of $\sigma_{\text{inel}} = 74.7 \pm 1.7$ mb \cite{Antchev:2013paa}.

Recently the ATLAS Collaboration presented a preliminary result on the measurement of $\sigma_{\text{inel}}$ at a centre-of-mass energy of $\sqrt{s}$ = 13 TeV \cite{ATLASXS13TeV}, and reported a measured  value of $65.2 \pm 0.8 \  \text{(exp.)} \pm 5.9 \  \text{(lum.)}$ mb within an acceptance of $\xi > 10^{-6}$ (corresponding to $M$ > 13 GeV) with $\xi = M^{2}/s$ and $M$ the mass of the largest diffractive dissociation system. This value was extrapolated to the total inelastic phase space, resulting in $\sigma_{\text{inel}} = 73.1 \pm 0.9 \  \text{(exp.)} \pm 6.6 \  \text{(lum.)} \pm 3.8 \  \text{(ext.)}$ mb.  

In these proceedings we present a measurement of the inelastic proton-proton cross section at $\sqrt{s}$ = 13 TeV with data collected by the CMS detector at the LHC \cite{CMS:2016ael}. The analysis method is based on the usage of extensive forward calorimetry available in CMS, and enables us to measure the inelastic cross section in two different detector acceptances. The first measurement only requires a signal in the Hadronic Forward (HF) calorimeters of CMS. They are placed at both sides of the interaction point and consist out of iron absorbers and quartz fibers for read out. The energy scale is known to an uncertainty of 10\%, and the used acceptance in this analysis is $3.152 < \left| \eta \right| < 5.205$. The second measurement includes the very forward CASTOR calorimeter of CMS. It consists out of tungsten absorbers and quarts plates for read out, and is only located at the minus side of the interaction point. The acceptance is $-6.6 < \eta < -5.2$, the energy scale uncertainty is 15\%, and the alignment of the detector is known to a $\pm 2$ mm accuracy \cite{CASTORDPS2016}. A detailed description of the CMS detector is available in \cite{Chatrchyan:2008aa}. The use of the very forward CASTOR calorimeter allows us to measure the inelastic cross section in the largest acceptance possible, and thus to reduce the extrapolation factor, and its uncertainty, towards the full inelastic phase space.

\section{Event selection and reconstruction}
The measurement uses LHC Run2 low pile-up proton-proton data with a centre-of-mass energy of $\sqrt{s}$ = 13 TeV with the solenoid of CMS both at B = 0 T and B = 3.8 T values. The CMS data acquisition was triggered by the presence of both beams in the interaction point, and this defines our total event sample (ZeroBias). Two different event selections are further applied offline to count the interactions in different detector acceptances:
\begin{itemize}
\item \textbf{HF OR}: require an energy deposit above 5 GeV in any of the two HF calorimeters.
\item \textbf{HF OR CASTOR}: require either an energy deposit above 5 GeV in any of the two HF calorimeters or an energy deposit above 5 GeV in the CASTOR calorimeter.
\end{itemize}
This first set of selected inelastic events is then corrected for remaining noise contributions:
\begin{equation}
N_{cor} = N_{ZB}\left[ \left( F_{ZB} - F_{EB} \right) + F_{EB} \left( F_{ZB} - F_{EB} \right) \right],
\end{equation}
where $N_{ZB}$ represents the number of ZeroBias triggered events; $F_{ZB}$ is the fraction of ZeroBias triggered events that are selected offline; and $F_{EB}$ is the fraction of no-beam triggered events that are selected offline. The selected number of interactions is further corrected for pile-up effects. The observed number of proton-proton collisions ($n$) per bunch crossing follows a Poisson distribution $P(n,\lambda)$ with average value $\lambda$. The probability to have no interaction in a ZeroBias sample is then given by $P(0,\lambda) \equiv \text{exp}(-\lambda)  = 1 - N_{cor} / N_{ZB}$, which allows one to determine the mean number of inelastic collisions per bunch crossing $\lambda = -\text{ln}(1 - N_{cor} / N_{ZB})$. In the data used this value ranges from 0.05 to 0.5. It is then possible to correct the inelastic event count with the following factor:
\begin{equation}
f_{\text{pu}} = \frac{\Sigma^{\infty}_{n=0} \  n P \left( n, \lambda \right) }{\Sigma^{\infty}_{n=1} \  P \left( n, \lambda \right)} = \frac{\lambda}{1 - P \left( 0, \lambda \right)}.
\end{equation}
This correction is applied bunch by bunch, and the total reconstructed number of interactions, corrected for contributions of noise and pile-up, is then given by:
\begin{equation}
N_{int} = \sum_{\text{bunches}} N^{b}_{cor} \times f^{b}_{\text{pu}},
\end{equation}
where $N_{cor}^{b}$ is the number of noise-corrected events per bunch, and $f_{\text{pu}}^{b}$ the pile-up correction factor per bunch. 

\section{Measurement of the visible inelastic cross section}
The measured number of inelastic interactions is corrected for various detector effects, such as the event selection efficiency and the resolution on the energy measurement, in order to compare to theoretical predictions. The acceptance on stable-particle level is defined as a function of $\xi$, which can be determined by first splitting the final state into systems X (negative side) and Y (positive side) separated by the largest rapidity gap, to then calculate their invariant masses $M_{X(Y)}$ and finally take the ratio with respect to the available centre-of-mass energy:
\begin{equation}
\xi_{X} = \frac{M^{2}_{X}}{s}, \quad  \xi_{Y} = \frac{M^{2}_{Y}}{s}, \quad \xi = \text{max}(\xi_{X},\xi_{Y}).
\end{equation}
Although the selection criteria on detector level are chosen to have the largest possible acceptance of inelastic events, low mass diffractive dissociation events will escape and not be included in the selected inelastic event sample. Hence only interactions with a value of $\xi_{X}$ or $\xi_{Y}$ above a certain threshold will be detected. These acceptance limits are obtained from a dedicated study using fully simulated events from various Monte Carlo event generators. The relation between the stable-particle level phase space definition and the offline detector level selection is quantified by efficiency and contamination factors. The efficiency ($\epsilon_{\xi}$) is defined as the fraction of selected stable-particle level events that also pass the offline detector-level selection, while the contamination ($b_{\xi}$) is defined as the fraction of offline detector-level selected events that are not part of the considered stable-particle level phase space domain. 

The optimal acceptances on stable-particle level are found to be $\xi > 10^{-6}$ for the HF OR detector-level selection, and the asymmetric (due to CASTOR being only present at the minus side) $\xi_{X} > 10^{-7} \text{or} \  \xi_{Y} > 10^{-6}$ for the HF OR CASTOR detector-level selection. Combining all ingredients, the stable-particle level inelastic cross section value for a given acceptance in $\xi$ is thus given by:
\begin{equation}
\sigma = \frac{N_{int} \left( 1 - b_{\xi} \right)}{\epsilon_{\xi} \int \mathcal{L} dt}.
\end{equation}
The systematic uncertainties are summarised in table \ref{tab:syst} and include the model dependence of the correction factors, the HF and CASTOR energy scale uncertainties, the alignment of the CASTOR calorimeter, and a run-to-run variation. Finally the uncertainty due to the luminosity measurement is given, which is known to a 2.7 \% (2.9 \%) precision for B = 3.8 T (0 T) data following a dedicated analysis of Van der Meer scans performed in August 2015 \cite{CMS:2016eto}.  

\begin{table}
\begin{center}
  \begin{tabular}{ l | c c }
       & $\sigma\left(\xi > 10^{-6}\right)$ (mb) & $\sigma\left(\xi_{X} > 10^{-7} \  \text{or} \  \xi_{Y} > 10^{-6} \right)$ (mb) \\  \hline
    Model dependence & 0.66 & 0.38 \\ 
    HF energy scale uncertainty & 0.34 & 0.13 \\
    CASTOR energy scale uncertainty & - & 0.04 \\
    CASTOR alignment & - & 0.03 \\
    Run-to-run variation & 0.15 & 0.14 \\ \hline
    Total & 0.76 & 0.44 \\ \hline
    Luminosity & 1.78 & 1.96 \\
  \end{tabular}
      \caption{Summary of systematic uncertainties from all sources.}
    \label{tab:syst}
\end{center}
\end{table}

\section{Results}
\begin{figure}[h]
        \centering
                \includegraphics[width=0.8\textwidth]{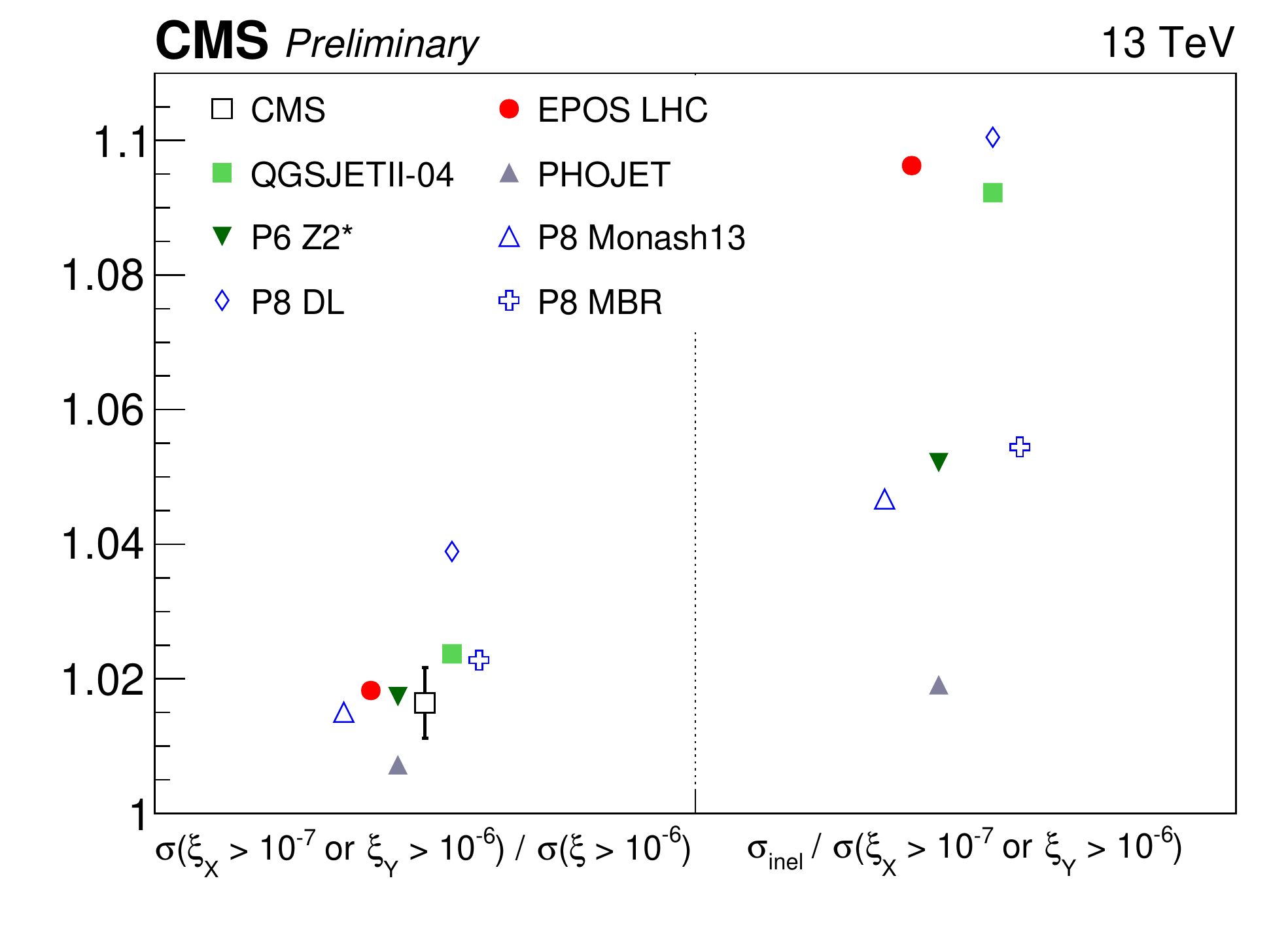}
                \caption{Ratio of the HF OR CASTOR ($\xi_{X} > 10^{-7} \  \text{or} \  \xi_{Y} > 10^{-6}$) measured cross section to the HF OR ($\xi > 10^{-6}$) acceptance (left bin); and the ratio of $\sigma_{\text{inel}}$ to the HF OR CASTOR ($\xi_{X} > 10^{-7} \  \text{or} \  \xi_{Y} > 10^{-6}$) cross section (right bin). The latter represents the extrapolation factors used to calculate $\sigma_{\text{inel}}$. \cite{CMS:2016ael}}
                \label{fig:FinalXSPlots_relative_bin1}
\end{figure}

\begin{figure}[h]
        \centering
                \includegraphics[width=0.8\textwidth]{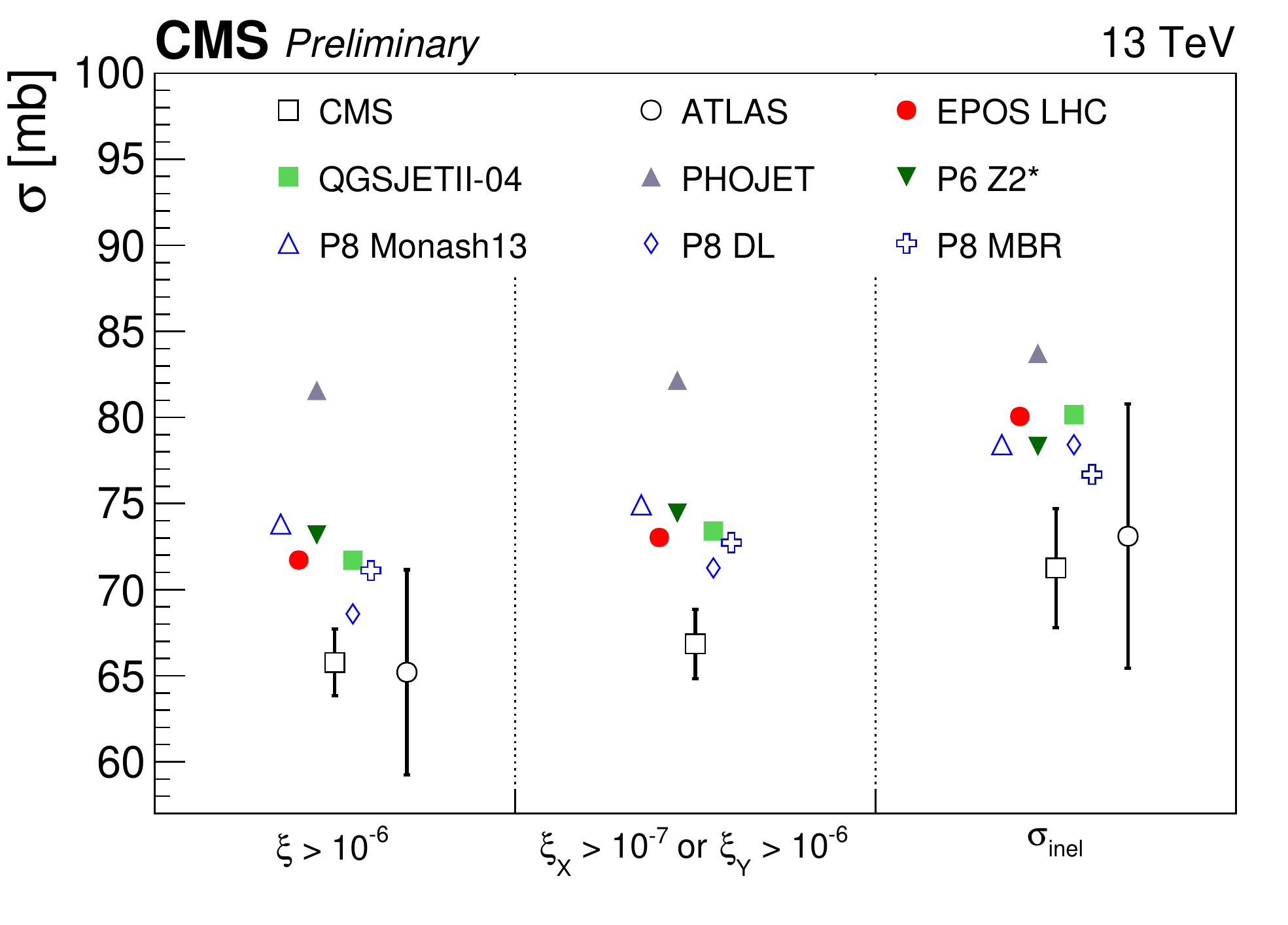}
        		\caption{Fully corrected inelastic cross sections measured in various phase space regions, compared to different models and the preliminary results of the ATLAS experiment. The data point for $\sigma_{\text{inel}}$ is calculated using a model-dependent extrapolation of the measured cross section for $\xi_{X} > 10^{-7} \  \text{or} \  \xi_{Y} > 10^{-6}$. \cite{CMS:2016ael}}
        		\label{fig:FinalXSPlots_absolute}
\end{figure}

The fully corrected inelastic cross section measured with the HF calorimeters only is:
\begin{equation}
\sigma \left( \xi > 10^{-6} \right) = 65.77 \pm 0.03 \  \text{(stat.)} \pm 0.76 \  \text{(sys.)} \pm 1.78 \  \text{(lum.) mb} ,
\end{equation}
and including the CASTOR calorimeter into the event selection to extend the acceptance yields:
\begin{equation}
\sigma \left( \xi_{X} > 10^{-7} \  \text{or} \  \xi_{Y} > 10^{-6} \right) = 66.85 \pm 0.06 \  \text{(stat.)} \pm 0.44 \  \text{(sys.)} \pm 1.96 \  \text{(lum.) mb.} 
\end{equation}
Figure \ref{fig:FinalXSPlots_relative_bin1} shows the ratio of the two measured cross sections in the left bin, indicating that most models are able to describe the relative increase from $\xi > 10^{-6}$ to $\xi_{X} > 10^{-7} \  \text{or} \  \xi_{Y} > 10^{-6}$. The right bin represents the model dependent extrapolation factors to go from the measured phase space to the full inelastic phase space domain. Using the average value of all models we obtain the following total inelastic cross section:
\begin{equation}
\sigma_{\text{inel}} = 71.26 \pm 0.06 \  \text{(stat.)} \pm 0.47 \  \text{(sys.)} \pm 2.09 \  \text{(lum.)} \pm 2.72 \  \text{(ext.) mb.} 
\end{equation}
The maximal difference between the model extrapolation factors is taken as uncertainty (ext.). Figure \ref{fig:FinalXSPlots_absolute} then shows the absolute values of all results, compared to various model predictions, and the preliminary ATLAS result \cite{ATLASXS13TeV}. While it is clear that the results of ATLAS and CMS are compatible within systematic uncertainties, all models predict an absolute cross section that is too high. 

\section{Summary}
A measurement of the inelastic proton-proton cross section at $\sqrt{s}$ = 13 TeV obtained with the CMS detector at the LHC has been presented. Visible cross sections in two acceptances are obtained: $\sigma \left( \xi > 10^{-6} \right) = 65.8 \pm 0.8 \  \text{(exp.)} \pm 1.8 \  \text{(lum.)}$ mb; and $\sigma \left( \xi_{X} > 10^{-7} \  \text{or} \  \xi_{Y} > 10^{-6} \right) = 66.9 \pm 0.4 \  \text{(exp.)} \pm 2.0 \  \text{(lum.)}$ mb. The latter visible cross section is extrapolated to the full inelastic phase space domain, yielding $71.3 \pm 0.5 \  \text{(exp.)} \pm 2.1 \  \text{(lum.)} \pm 2.7 \  \text{(ext.)}$ mb. The measured cross section is lower than predicted by models for hadronic scattering.

%\section{Acknowledgements}
%We are grateful to the organisers, convenors and participants of the DIS conference for this opportunity to present and discuss our results in the Small-x and Diffraction working group.

\end{document}